\documentclass[prl,showpacs,showkeys,nofootinbib,twocolumn,floatfix]{revtex4}

\usepackage{graphicx}  
\usepackage{amsmath}
\usepackage{bm}  
\usepackage{ulem}
\usepackage{amssymb}
\usepackage{comment}
\usepackage{lineno}

\newcommand{\bea}{\begin{eqnarray}}
\newcommand{\eea}{\end{eqnarray}}
\newcommand{\beq}{\begin{equation}}
\newcommand{\eeq}{\end{equation}}
\newcommand{\bqa}{\begin{eqnarray}}
\newcommand{\eqa}{\end{eqnarray}}

\begin{document}

\title{
Triangle Singularity in the Production of $\bm{X(3872)}$  \\
and a Photon in $\bm{e^+e^-}$ Annihilation
}

\author{Eric Braaten}
\email{braaten.1@osu.edu}
\affiliation{Department of Physics,
         The Ohio State University, Columbus, OH\ 43210, USA}

\author{Li-Ping He}
\email{he.1011@buckeyemail.osu.edu}
\affiliation{Department of Physics,
         The Ohio State University, Columbus, OH\ 43210, USA}

\author{Kevin Ingles}
\email{ingles.27@buckeyemail.osu.edu}
\affiliation{Department of Physics,
         The Ohio State University, Columbus, OH\ 43210, USA}

\date{\today}

\begin{abstract}
If the $X(3872)$ is a weakly bound charm-meson molecule, 
it can be produced in $e^+ e^-$ annihilation by the creation of $D^{*0} \bar D^{*0}$  from a virtual photon
followed by the  rescattering of the charm-meson pair into $X$ and a photon.
A triangle singularity produces a narrow peak in the cross section for $e^+ e^- \to X \gamma$ about 2.2~MeV
above the $D^{*0} \bar{D}^{*0}$ threshold. 
We predict the  normalized cross section in the region near the peak.
The peak from the triangle singularity may be observable by the BESIII detector.

\end{abstract}

\smallskip
\maketitle

Since early in this century, a large number of exotic hadrons whose constituents include a heavy quark and its antiquark
have been discovered in high energy physics experiments 
\cite{Chen:2016qju,Hosaka:2016pey,Lebed:2016hpi,Esposito:2016noz,Guo:2017jvc,Ali:2017jda,Olsen:2017bmm,Karliner:2017qhf,Yuan:2018inv}. 
The first on the  list of these exotic heavy hadrons is the $X(3872)$ meson,
which was discovered  in 2003  in exclusive decays of $B^\pm$ mesons into $K^\pm X$ 
by observing the decay of $X$ into $J/\psi\, \pi^+\pi^-$ \cite{Choi:2003ue}. 
The $J^{PC}$ quantum numbers of $X$ were eventually determined to be $1^{++}$ \cite{Aaij:2013zoa}.
Its mass  is extremely close to the $D^{*0} \bar D^0$  threshold,
with the difference being only $0.01 \pm 0.18$~MeV \cite{Tanabashi:2018oca}.
This suggests that $X(3872)$ is a weakly bound S-wave charm-meson molecule
with the flavor structure
\begin{equation}
\big| X(3872) \rangle = \frac{1}{\sqrt2}
\Big( \big| D^{*0} \bar D^0 \big\rangle +  \big| D^0 \bar D^{*0}  \big\rangle \Big).
\label{Xflavor}
\end{equation}
There are however alternative models for the $X$
\cite{Chen:2016qju,Hosaka:2016pey,Lebed:2016hpi,Esposito:2016noz,Guo:2017jvc,Ali:2017jda,Olsen:2017bmm,Karliner:2017qhf}.
The $X$ has been observed in many more decay modes than any of the other exotic heavy hadrons.
In addition to  $J/\psi\, \pi^+\pi^-$, it has been observed in 
 $J/\psi\, \pi^+\pi^-\pi^0$, $J/\psi\, \gamma$, $\psi(2S)\, \gamma$, $D^0 \bar D^0 \pi^0$, $D^0 \bar D^0 \gamma$, 
 and most recently $\chi_{c1}\, \pi^0$ \cite{Ablikim:2019soz}.
Despite its observation  in  7 different decay modes,  a consensus on the nature of $X$ has not been achieved.

There may be aspects of the production of $X$ that are more effective at discriminating
between models than the decays of $X$.
One way in which the nature of a hadron can be revealed by its production is through {\it triangle singularities}.
Triangle singularities are kinematic singularities that  arise if three virtual particles that form a triangle 
in a Feynman diagram can all be on their mass shells simultaneously.
There have been several previous investigations of the effects of triangle singularities on the production of
exotic heavy mesons \cite{Szczepaniak:2015eza,Liu:2015taa,Szczepaniak:2015hya,Guo:2017wzr}.
Guo has recently pointed out that  any high-energy process 
that can create $D^{*0} \bar{D}^{*0}$ at short distances in an S-wave channel
will produce $X \gamma$ with a narrow peak near the $D^{*0} \bar{D}^{*0}$ threshold 
due to a triangle singularity \cite{Guo:2019qcn}. 
One such process is electron-positron annihilation,
which can create an S-wave $D^{*0} \bar{D}^{*0}$ pair recoiling against a $\pi^0$.
Guo suggested that the peak in the   line shape for $X  \gamma$ due to the  triangle singularity
could be used to  determine the binding energy  of $X$ more accurately than a direct mass measurement.

If the $X$ is a weakly bound charm-meson molecule, it
can be produced by any reaction that can produce its constituents $D^{*0} \bar D^0$ and $D^0 \bar D^{*0}$. 
It can be produced by the creation of $D^{*0} \bar D^0$ and $D^0 \bar D^{*0}$
at short distances of order $1/m_c$, where $m_c$ is the charm quark mass,
followed by the binding of the charm mesons into $X$ at longer distances.
The $X$ can also be produced by the  creation of $D^* \bar{D}^*$ at short distances followed by the rescattering of 
the charm-meson pair into $X$ and a pion at longer distances \cite{Braaten:2018eov}.
The $D^* \bar{D}^*$ rescattering mechanism predicts that the Dalitz plot from the decay of a $B$ meson into $K X \pi$ 
should be dominated by a resonance band from the decay  into $K^*(892)\,  X$ and a smooth distribution 
in $X\pi$ invariant mass from the rescattering of $D^* \bar{D}^*$ \cite{Braaten:2019yua}.
The production of $X$ accompanied by a pion from $D^* \bar{D}^*$ rescattering 
provides an additional production mechanism for $X$ at a high-energy hadron collider 
that  could help explain the large prompt production rate of $X$ 
that has been observed at the Tevatron and the LHC  \cite{Braaten:2018eov,Braaten:2019sxh}.

The quantum numbers $1^{++}$ of the $X$ imply that $X\gamma$ can be produced directly 
by $e^+ e^-$ annihilation into a virtual photon.
This process can  proceed through the creation of $D^{*0} \bar{D}^{*0}$ at short distances in a P-wave channel, 
followed by the rescattering of the charm-meson pair into $X \gamma$.
The production of $X \gamma$ in $e^+ e^-$ annihilation near the $D^{*0} \bar D^{*0}$ threshold
was discussed previously by Dubynskiy and Voloshin  \cite{Dubynskiy:2006cj}.
They calculated the absorptive contribution to the cross section from 
$e^+ e^-$ annihilation into on-shell charm mesons $D^{*0} \bar{D}^{*0}$ 
followed by their rescattering into $X \gamma$.
They predicted that the  cross section has a peak only a few MeV above the $D^{*0} \bar D^{*0}$ threshold,
although they did not predict its normalization.
In retrospect, this peak comes from a triangle singularity.

In this paper, we calculate the cross section for   $e^+e^-\to X \gamma$ near the  $D^{*0} \bar D^{*0}$ threshold.
We show that the absorptive contribution considered in Ref.~\cite{Dubynskiy:2006cj}
is not a  good approximation for the cross section.
We give a normalized prediction for the cross section
by using results from a fit to Belle data on  $e^+e^-\to D^*\bar D^*$ by Uglov {\it et al.}\    \cite{Uglov:2016orr}. 
The peak from the triangle singularity is large enough that it could be observable by the BESIII detector.

\begin{figure}[ht]
\includegraphics*[width=0.45\linewidth]{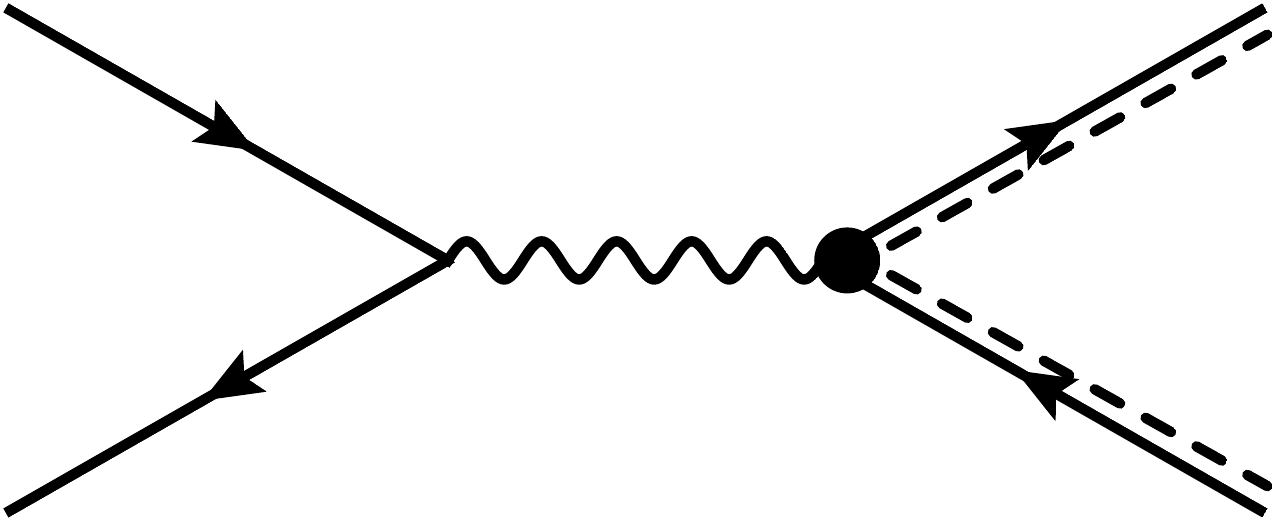} 
\caption{Feynman diagram for $e^+e^-\to  D^{*0} \bar D^{*0}$. 
The spin-1 charm mesons $D^{*0}$ and $\bar D^{*0}$ are represented by double lines 
consisting of a dashed line and a solid line with an arrow.
}
\label{fig:eetoD*D*}
\end{figure}

A pair of spin-1 charm mesons $D^{*0} \bar D^{*0}$ can be produced from the annihilation of $e^+e^-$ 
into a virtual photon. The Feynman diagram for this process is shown in Fig.~\ref{fig:eetoD*D*}.
We use nonrelativistic normalizations for the  charm mesons in the final state.
In the center-of-momentum frame, the matrix element has the form
\begin{equation}
\mathcal{M}= -i\frac{e^2}{s}\, \bar v \gamma^i u\,J^i ,
\label{MeetoD*D*}
\end{equation}
where $s$ is the square of the center-of-mass energy, $\bar v$ and $u$ are the spinors for the colliding
$e^+$ and $e^-$, and $\bm{J}$ is the matrix element of the electromagnetic current
between the QCD vacuum and the $D^{*0} \bar D^{*0}$ state.
Near the threshold for producing $D^{*0} \bar D^{*0}$, the charm-meson pair must be produced in a P-wave state 
with total spin 0 or 2.  The matrix element of the current that creates  $D^{*0}$ and $\bar D^{*0}$ 
with momenta $+\bm{k}$ and $-\bm{k}$ and with polarization vectors $\bm{\varepsilon}$ and $\bar{\bm{\varepsilon}}$  
can be expressed as 
$J^i=\mathcal{A}^{ijkl} k^j\varepsilon^{*k}\bar{\varepsilon}^{*l}$.
The Cartesian tensor $\mathcal{A}^{ijkl}$ is
\begin{equation}
\mathcal{A}^{ijkl}= A_0 \, \delta^{ij} \delta^{kl} 
+ \frac{3}{2\sqrt{5}} A_2 \left(\delta^{ik} \delta^{jl}+\delta^{il} \delta^{jk} - \frac23 \delta^{ij} \delta^{kl}  \right) ,
\label{A[eetoD*D*]}
\end{equation}
where  $A_0$ and $A_2$ are amplitudes for creating $D^{*0} \bar D^{*0}$ with total spin 0 and 2, respectively. 
The cross section for $e^+e^-$ annihilation into $D^{*0} \bar D^{*0}$ near the  threshold is
\begin{equation}
\sigma[ e^+ e^- \to D^{*0} \bar D^{*0}] = \frac{4\pi \alpha^2 M_{*0}}{s^2}
\left[ |A_0|^2  + |A_2|^2  \right] k^3  ,
\label{sigma}
\end{equation}
where  $M_{*0}$ is the mass of the  $D^{*0}$
and $k =[ M_{*0} (\sqrt{s} - 2 M_{*0} )]^{1/2}$ is the relative momentum of the $D^{*0} \bar D^{*0}$ pair.

The Belle collaboration has measured exclusive cross sections for $e^+ e^-$ annihilation 
into several pairs of charm mesons, including $D^{*+} D^{*-}$ \cite{Abe:2006fj,Pakhlova:2008zza}.
Uglov {\it et al.}\  have analyzed the Belle data  using a unitary approach 
based on a coupled channel model  \cite{Uglov:2016orr}. 
They included a  spin-2 F-wave amplitude for $e^+ e^- \to D^* \bar D^*$ as well as 
spin-0 and spin-2 P-wave amplitudes.  Near the $D^{*+} D^{*-}$ threshold at 4020.5~MeV,
the  spin-0 and spin-2 P-wave contributions to the cross sections have the $k^3$ behavior in Eq.~\eqref{sigma}.
A fit to the two terms in the cross section in Eq.~\eqref{sigma}, 
with $M_{*0}$ replaced by the mass $M_{*1}$ of the $D^{*+}$ and $k =[ M_{*1} (\sqrt{s} - 2 M_{*1} )]^{1/2}$, gives
\begin{equation}
|A_0| = 8~\mathrm{GeV}^{-1}, \qquad |A_2| = 15~\mathrm{GeV}^{-1} .
\label{A0,A2}
\end{equation}
These coefficients have natural magnitudes of order $1/m_\pi$.
The factor $|A_0|^2 + |A_2|^2$ in Eq.~\eqref{sigma} is determined more accurately by fitting cross sections 
than the ratio $|A_2|/|A_0|$.

The values of $|A_0|$  and $|A_2|$ for $e^+ e^- \to D^{*+} D^{*-}$ in Eq.~\eqref{A0,A2} 
can be inserted into Eq.~\eqref{sigma} to predict the cross section for $e^+ e^- \to D^{*0} \bar D^{*0}$ 
near  the $D^{*0} \bar D^{*0}$ threshold at 4013.7~MeV. 
This prediction is based on the reasonable assumption that the creation of the charm-meson 
pair  proceeds through the direct coupling of the virtual photon to the charm quark 
and that the contribution from its direct coupling to the light quark is negligible.

If the $X(3872)$ is a weakly bound charm-meson molecule, its constituents are
the superposition of charm mesons in Eq.~\eqref{Xflavor}.
The reduced mass of $D^{*0} \bar D^0$ is $\mu=M_{*0}M_0/(M_{*0}\!+\!M_0)$,
where $M_0$ is the mass of the $D^0$.
The mass difference between the $D^{*0}$ and $D^0$ is $\delta = M_{*0}\!-\!M_0= 142.0$~MeV.
The present value of the difference $E_X$ between the mass of the $X$ 
and the energy of the $D^{*0} \bar D^0$ scattering threshold is \cite{Tanabashi:2018oca}
\begin{equation}
E_X \equiv M_X - (M_{*0}\!+\!M_0) =( +0.01 \pm 0.18)~\mathrm{MeV}.
\label{deltaMX}
\end{equation}
If this difference  is negative,  $X$ is a bound state with binding energy $|E_X|$ and
 binding momentum $\gamma_X = \sqrt{2 \mu|E_X|}$.
The central value in Eq.~\eqref{deltaMX} corresponds to a charm-meson pair above the scattering threshold.
The value lower by $1\sigma$ corresponds to a bound state with binding energy $|E_X| =0.17$~MeV
and binding momentum $\gamma_X=18$~MeV.

\begin{figure}[ht]
\includegraphics*[width=0.45\linewidth]{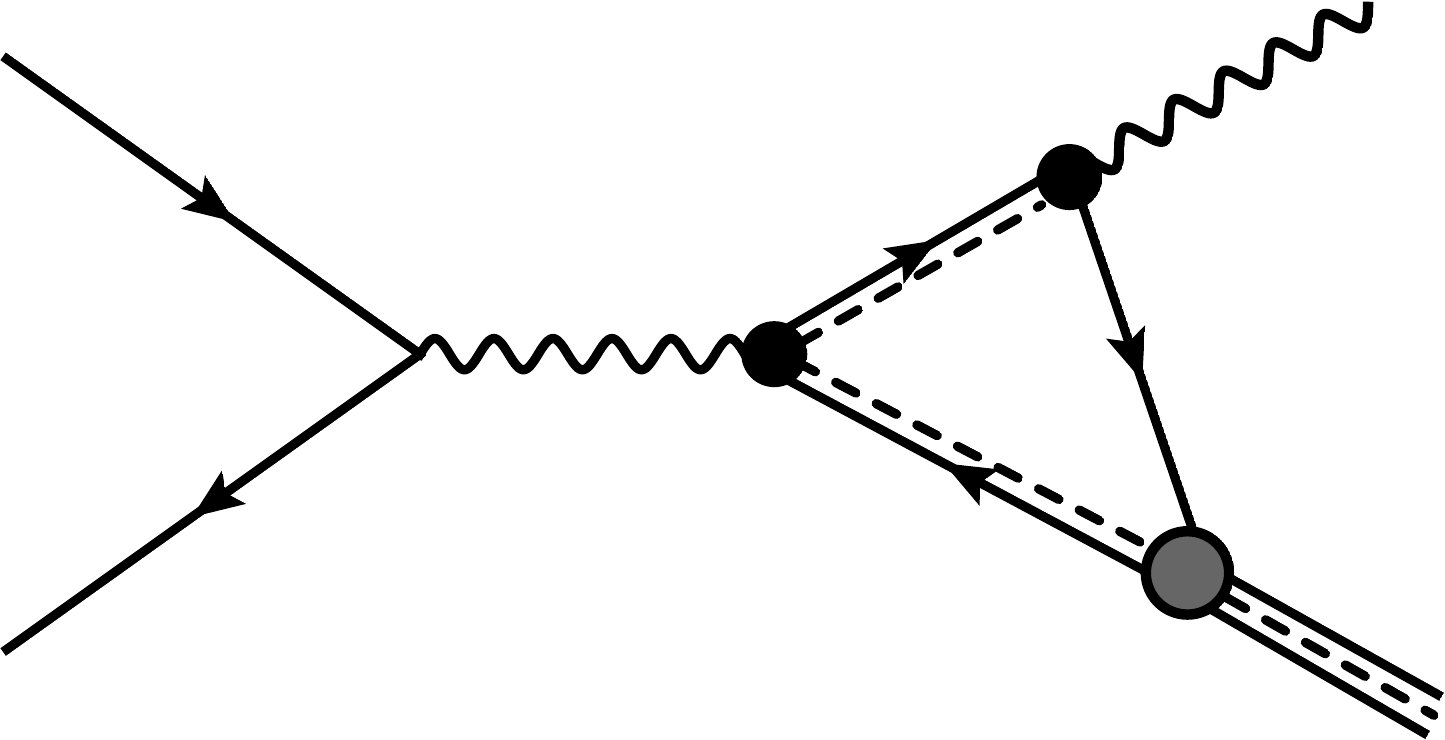} \quad
\includegraphics*[width=0.45\linewidth]{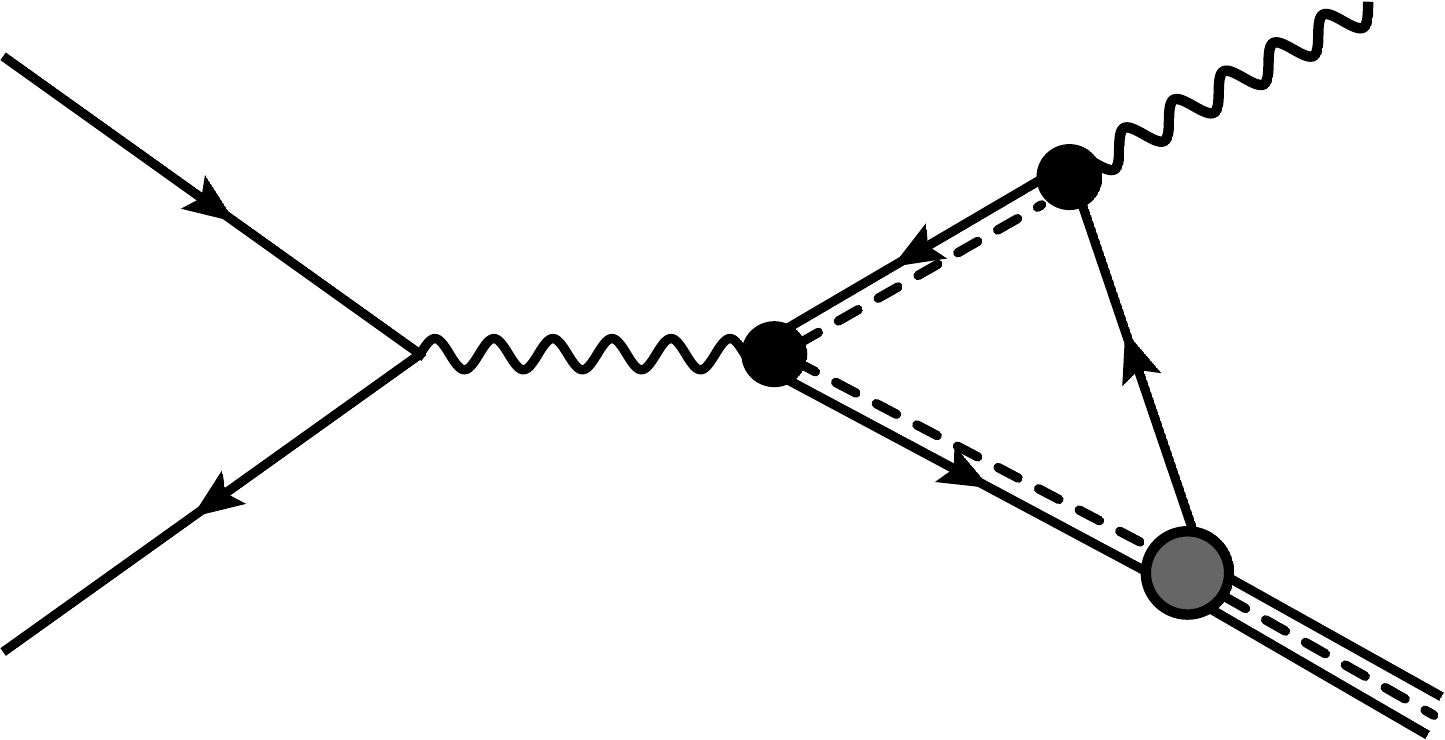} 
\caption{Feynman diagrams for $e^+e^-\to \gamma X$ from rescattering of  $D^{*0} \bar D^{*0}$. 
The $X$ is represented by a triple line consisting of two solid lines and a dashed line. 
The spin-0 charm mesons $D^0$ and $\bar D^0$ are represented by solid lines with an arrow.
}
\label{fig:eetogammaX}
\end{figure}

The $X$ can be produced in $e^+e^-$ annihilation through the creation of $D^{*0} \bar D^{*0}$
by a virtual photon followed by the rescattering of the charm-meson pair into $X\gamma$.
The Feynman diagrams for this process are shown in Fig.~\ref{fig:eetogammaX}.
The vertex for the virtual photon to create $D^{*0}$ and $\bar D^{*0}$ 
with momenta $+\bm{k}$ and $-\bm{k}$ and  vector indices $m$ and $n$ is $e \mathcal{A}^{ijmn} k^j$,
where the Cartesian tensor is given in Eq.~\eqref{A[eetoD*D*]}.
The vertex for the transition of $D^{*0}$ to $D^0\gamma$ with a photon of momentum  $\bm k$
is $-e\nu \epsilon^{ijm} k^m$, where $i$  is the vector index of $D^{*0}$. 
The transition magnetic moment $e\nu$ can be determined from the radiative decay width of $D^{*0}$ \cite{Rosner:2013sha}:
$\nu= 0.95~\mathrm{GeV}^{-1}$.  
The binding of $D^{*0} \bar D^{0}$ or $D^{0} \bar D^{*0}$ into $X$ 
can be described within an effective field theory called XEFT \cite{Fleming:2007rp,Braaten:2015tga}.
The vertices for the couplings of $D^{*0}\bar D^0$ to $X$ and $D^{0}\bar D^{*0}$ to $X$ can be expressed as 
$i (\sqrt{\pi \gamma_X}/\mu)\, \delta^{kl}$,
where $\gamma_X$ is the binding momentum of the $X$ and
$k$ and $l$ are the vector indices of the  spin-1 charm meson and the $X$ \cite{Braaten:2010mg}.
 
The matrix element for $e^+e^-\to X \gamma$ is the sum of the two diagrams  in Fig.~\ref{fig:eetogammaX}.  
We use nonrelativistic propagators for the charm mesons.
The matrix element for producing $X$ and $\gamma$ with momenta $\bm q$ and $-\bm q$
and with polarization vectors $\bm \varepsilon_X$ and $\bm \varepsilon_\gamma$ can be expressed as
\begin{equation}
\mathcal{M} = \frac{2e^3 \nu M_{*0}M_0}{s\, \mu} \, \bar v \gamma^i u \, \mathcal{J}^i\, F(W).
\label{MeetoXgamma}
\end{equation}
The current $\mathcal{J}^i$ is
\begin{equation}
\bm{\mathcal{J}} \!= \!
\bigg(A_0 - \frac{1}{\sqrt{5}} A_2 \bigg)  (\hat{\bm q} \times \bm \epsilon_\gamma \cdot \bm \epsilon_X) \hat{\bm q}
+\frac{3}{2\sqrt{5}} A_2 (\hat{\bm q} \cdot \bm \epsilon_X)  \hat{\bm q} \times \bm \epsilon_\gamma .
\label{currentJ}
\end{equation}
The scalar loop amplitude $F(W)$ is a function of
the center-of-mass energy $W= \sqrt{s} - 2 M_{*0}$ relative to the $D^{*0} \bar D^{*0}$ threshold.
It can be calculated analytically by integrating over the loop energy using a 
contour integral around the pole of one propagator, combining the remaining 
two propagators with a Feynman parameter $x$, integrating over the loop 
momentum, and finally integrating over $x$:
\begin{eqnarray}
F(W) &=& - i\frac{\mu \sqrt{\pi \gamma_X}}{4\pi M_0} q
 \bigg(\frac{B^2 - A ^2- C^2}{2 C^2}    \log\frac{A +B + C}{A+B-C}
 \nonumber\\
 &&\hspace{2cm}
 +\frac{A  - B}{C}  \bigg) ,
 \label{Fanalytic}
\end{eqnarray}
where
\begin{subequations}
\begin{eqnarray}
A&=&   \sqrt{k^2+i M_{*0}\Gamma_{*0} },
    \\
B &=&  i\sqrt{\gamma_X^2- i \mu \Gamma_{*0}} ,
\label{coeffa}
    \\
C&=&  (\mu/M_0)  q ,
\end{eqnarray}
\label{coefficientsabc}%
\end{subequations}
and $\Gamma_{*0} = 56$~keV is the predicted width of the $D^{*0}$ \cite{Rosner:2013sha}.
The dependence on $W$ is through $k^2=M_{*0}W$  and 
the momentum $q$ of $X$, which is equal to the energy of the photon
and is determined by energy conservation:
\begin{equation}
W = q - \delta + \frac{q^2}{2M_X} - \frac{\gamma_X^2}{2\mu} .
\label{q-W}
\end{equation}

The loop amplitude in Eq.~\eqref{Fanalytic}  has a  triangle singularity
 in the limit where the binding energy of $X$ is 0 and the decay width of the $D^{*0}$ is 0. 
The singularity arises from  the integration region where the three charm mesons whose  lines form triangles
in the diagrams in Fig.~\ref{fig:eetogammaX} are all on their mass shells simultaneously. 
The two charm mesons that become constituents of the $X$ are both  on their mass shells 
in the limit where the binding energy is 0.  There is a specific energy $W_\triangle$ where
the spin-1 charm meson that  emits the photon can also be on its mass shell.
The triangle singularity is a  logarithmic divergence of Eq.~\eqref{Fanalytic}
 in the limits $\gamma_X \to 0$ and $\Gamma_{*0} \to 0$.
In these limits, the denominator of the argument of the logarithm is zero at the energy for which $k = (\mu/M_0)q$.
The energy $W_\triangle$  can be obtained by solving this equation for $W$ using  Eq.~\eqref{q-W} for $q$.  
A simple approximation to the solution  for $W$ is
\begin{equation}
W_\triangle = (\mu/M_0)^2\delta^2/M_{*0} .
    \label{Wtriangle}
\end{equation}
The predicted energy is $W_\triangle=2.7$~MeV.

The cross section for $e^+ e^-$ annihilation into $X \gamma$ 
 at center-of-mass energy near the $D^{*0}\bar D^{*0}$ threshold is
\begin{eqnarray}
\sigma[e^+ e^- \to X \gamma] &=&
\frac{128 \pi^2\alpha^3 \nu^2 M_X^2}{3 s^2  (1+q/M_X)} 
\nonumber\\
&&\hspace{-2.0cm}\times
\bigg(\bigg| A_0 - \frac{1}{\sqrt{5}} A_2\bigg|^2  + \frac{9}{20} |A_2|^2 \bigg)\,
q \big| F(W) \big|^2.
\label{sigmagammaXana}
\end{eqnarray}
We have used nonrelativistic phase space for $X$ and relativistic phase space for the photon.
The factor that depends on $A_0$  and $A_2$ differs from $|A_0|^2 + |A_2|^2$
by a multiplicative factor that depends on $A_2/A_0$ and can range from 0.34 to 1.31. 

\begin{figure}[t]
\includegraphics*[width=1.0\linewidth]{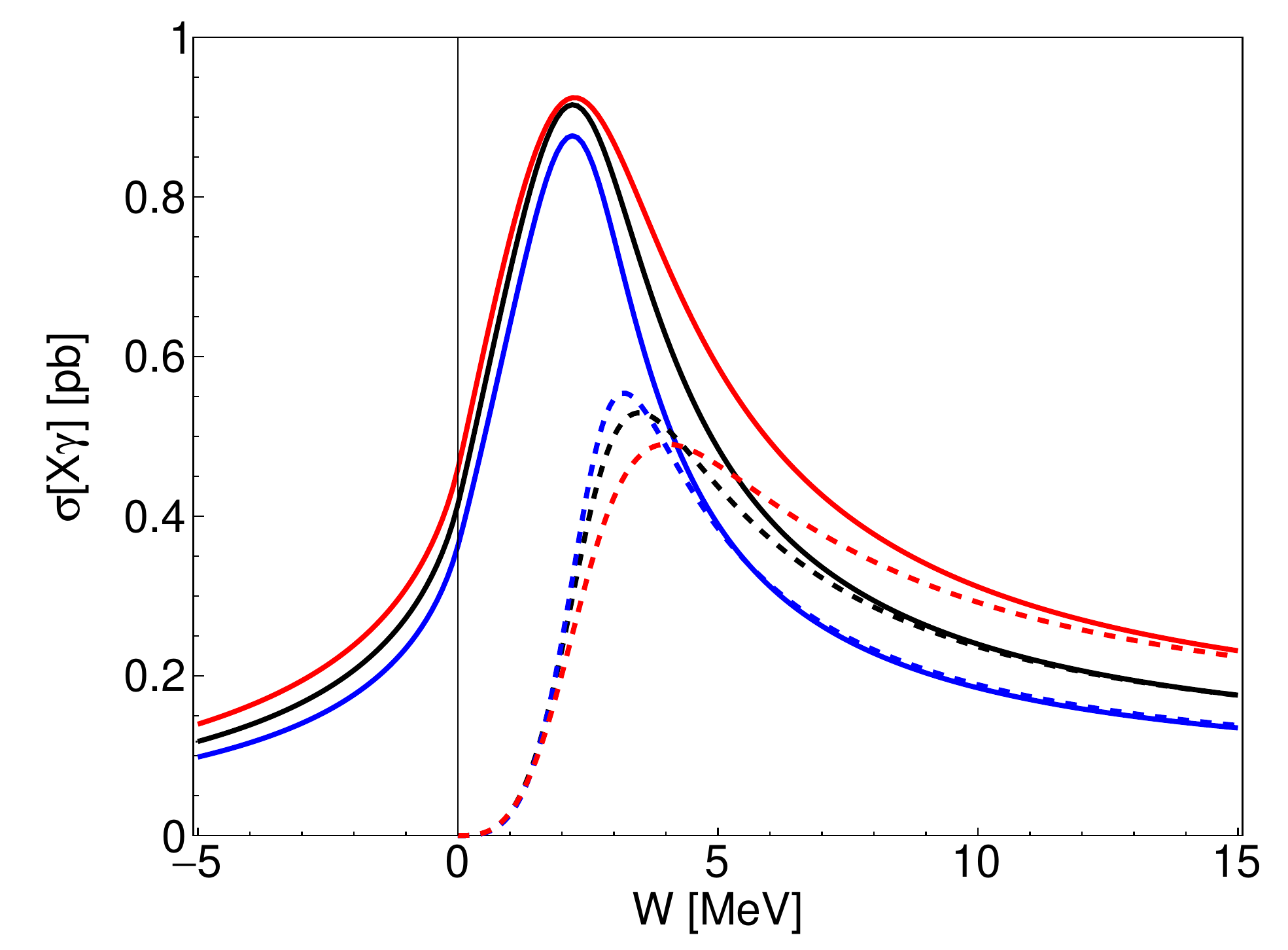} 
\caption{
Cross section for $e^+ e^- \to X(3872)\,  \gamma$ as a function of the 
center-of-mass energy $W$ relative to the $D^{*0} \bar D^{*0}$ threshold.
The solid curves in order of decreasing cross sections are for binding energies 
$|E_X| = 0.30$~MeV, 0.17~MeV, and 0.10~MeV.
The dashed curves are the absorptive contributions,
which approach the corresponding cross sections as $W$ increases.
}
\label{fig:sigma-absorp}
\end{figure}

The cross section for $e^+e^- \to X \gamma$  in Eq.~\eqref{sigmagammaXana}
near the $D^{*0} \bar D^{*0}$ threshold  is shown  in Fig.~\ref{fig:sigma-absorp} for three values of the
binding energy: $|E_X| = 0.30$~MeV, 0.17~MeV, and 0.10~MeV.
For smaller values of $E_X$, the line shape becomes sensitive to the unknown decay width of $X$.
It is necessary to take into account the imaginary part of $\gamma_X$, and it becomes increasingly difficult 
to accomodate the significant branching fraction into $J/\psi\, \pi^+ \pi^-$ \cite{Braaten:2007dw}.
We have chosen the value of $A_2/A_0$ that maximizes the normalization factor in the cross section
given the value $|A_0|^2 + |A_2|^2 = 280~\mathrm{GeV}^{-2}$ from Eq.~\eqref{A0,A2}.
The minimum normalization factor is smaller by a factor of 3.8.
The cross section has a narrow peak produced by the triangle singularity. 
The position of the peak, which is insensitive to the binding energy, is  2.2~MeV 
above the $D^{*0} \bar D^{*0}$ threshold, near the prediction from Eq.~\eqref{Wtriangle}.
The height of the peak is also insensitive to $|E_X|$.  
The full width at half maximum  is 5.1~MeV  at $|E_X|=0.17$~MeV, and it decreases as $|E_X|$ decreases.
 Beyond the peak, the cross section decreases to a local minimum at an energy $W$ near 40~MeV,
 before  increasing because of the $k^3$ dependence of the P-wave cross section for producing $D^{*0}\bar D^{*0}$.
 The minimum  is 0.1~pb for $|E_X| = 0.17$~MeV. 

 The loop amplitude $F(W)$ in Eq.~\eqref{Fanalytic} has an absorptive contribution that corresponds to 
$e^+e^-$ annihilation into
on-shell charm mesons $D^{*0} \bar D^{*0}$ followed by the rescattering of the charm mesons into $X\gamma$.
The  absorptive contribution to $F(W)$ is the imaginary part of  Eq.~\eqref{Fanalytic} in the limit $\Gamma_{*0}\to 0$:
\begin{eqnarray}
\mathrm{Im} \big[F(W)\big] &=& \frac{\sqrt{\pi \gamma_X}}{4\pi} k
\bigg( \frac{ (\mu/M_0)^2 q^2 + k^2 + \gamma_X^2}{4(\mu/M_0) qk}  
\nonumber\\
&& \times
\log\frac{[(\mu/M_0)q + k]^2 + \gamma_X^2}{[(\mu/M_0)q - k]^2 + \gamma_X^2} - 1 \bigg).
\label{ImFabs-analytic}
\end{eqnarray}
The absorptive contribution to the cross section, which is obtained by replacing $F(W)$ in 
Eq.~\eqref{sigmagammaXana} by $\mathrm{Im} [F(W)]$ in Eq.~\eqref{ImFabs-analytic},
is shown in Fig.~\ref{fig:sigma-absorp} for three values of $|E_X|$.
The absorptive contribution is zero below the $D^{*0} \bar D^{*0}$ threshold.
For $|E_X|=0.17$~MeV, the position of the peak in the absorptive contribution is 1.3~MeV higher
than that of the full cross section.  The height of the peak is 58\% of that of the full cross section.  
Thus the absorptive contribution is not a good approximation to the cross section
in the triangle singularity region.
At larger energies, the absorptive contribution quickly approaches the full cross section.
 
In the limit $\Gamma_{*0} \to 0$, the loop amplitude can be expressed as the momentum integral 
\begin{eqnarray}
 F(W)&=& 
- \frac{1}{\sqrt{2}\,M_X}
 \int\!\! \frac{d^3k}{(2\pi)^3} \, 
   \frac{(\bm q \cdot \bm k) \, \psi_X(| \bm k -(\mu/M_0)\bm q|)}{W  - \bm k^2/M_{*0}+ i \epsilon},
   \nonumber\\
\label{eetoXgamma-wfX}
\end{eqnarray}
where $\psi_X(k)$ is the universal momentum-space Schr\"odinger wavefunction
 for a weakly-bound S-wave molecule normalized so $\int d^3k/(2 \pi)^3\, |\psi(k)|^2 = 1$.  
 In the rest frame, this wavefunction as a function of the relative momentum $k$ is
\begin{equation}
\psi_X(k) = \frac{ \sqrt{8 \pi \gamma_X}}{k^2 + \gamma_X^2}.
\label{psiX-k}
\end{equation}
The wavefunction in the integrand of Eq.~\eqref{eetoXgamma-wfX}  is  in a moving frame 
where the bound state has momentum $\bm q$. 
It has a maximum as  function of $\bm k$ when the two constituents have equal velocities.

In Ref.~\cite{Dubynskiy:2006cj}, Dubynskiy and Voloshin (DV) calculated
the absorptive contribution to the cross section for $e^+e^- \to  X \gamma$
using a wavefunction that reduces in the rest frame to
\begin{equation}
\psi_\mathrm{DV}(k) = 
\frac{\sqrt{ 8 \pi \gamma_X\Lambda(\Lambda \!+\!\gamma_X)}}{\Lambda-\gamma_X} 
\bigg(\frac{1}{k^2 + \gamma_X^2} - \frac{1}{k^2 + \Lambda^2} \bigg),
\label{psiDV-k}
\end{equation}
where $\Lambda$ is an adjustable parameter.
The DV wavefunction was used previously by Voloshin in a study of the decays of $X(3872)$ into 
$D^0 \bar D^0 \gamma$ \cite{Voloshin:2005rt}.  
Voloshin identified the parameter $\gamma_X$ in Eq.~\eqref{psiDV-k} with the binding momentum of $X$.
This identification can be justified rigorously only in the limit $\Lambda \to \infty$ in which
the wavefunction reduces to that in Eq.~\eqref{psiX-k}.
The subtraction in the DV wavefunction in Eq.~\eqref{psiDV-k} makes it decrease as $1/k^4$ 
when $k$ is much larger than the momentum scale $\Lambda$. 
In Ref.~\cite{Dubynskiy:2006cj}, DV illustrated their results 
using the values $\Lambda = 200$~MeV and $\Lambda=300$~MeV.
For the wavefunction in the moving frame, 
DV used $\psi_\mathrm{DV}(|\bm{k} - \tfrac12 \bm{q}| )$.
It has a maximum as  function of $\bm k$ when the two constituents have equal momenta.
Since $\mu/M_0=0.518$ is close to $\tfrac12$, the different prescription for the wavefunction 
in the moving frame has a small effect on the cross section.
The subtraction in Eq.~\eqref{psiDV-k} also has a small effect in the triangle singularity region.

The BESIII collaboration has measured the cross section for $e^+ e^-$ annihilation into $X\gamma$
at center-of-mass energies ranging from 4.008~GeV to 4.6~GeV 
by observing $X$ in the final states $J/\psi\, \pi^+\pi^-$ and $J/\psi\, \omega$ \cite{Ablikim:2013dyn,Ablikim:2019zio}.
They did not measure the cross section at energies  between  4.009~GeV and 4.178~GeV,
which includes the energy 4.016~GeV of the peak from the triangle singularity.
The BESIII collaboration measured the cross section in 10~MeV steps 
between 4.178~GeV and 4.278~GeV  \cite{Ablikim:2019zio}. 
The largest value of the product $\mathrm{Br}\, \sigma$ of the cross section
and the branching fraction into  $J/\psi\, \pi^+\pi^-$ was about 0.5~pb.
Upper and lower bounds can be put on the branching fraction Br for the decay of $X$ into  $J/\psi\, \pi^+\pi^-$:
$4\% < \mathrm{Br} < 33\%$ \cite{Braaten:2019ags}.
Thus the height of the peak from the triangle singularity at 4.016~GeV could be 
 a significant fraction of the cross section measured in this higher energy region.

Our predictions for the $X+\gamma$ peak from the triangle singularity
are based on the assumption that the $X$ is a weakly-bound charm 
meson molecule whose decay width is smaller than its binding energy.
If the decay width is larger than the binding energy or if the $X$ is 
instead a virtual state, our assumption that it couples to the charm mesons 
through a momentum-independent vertex breaks down.  This could have  
a significant effect on the peak.  The peak could also be modified if there
is a multiple fine tuning of hadronic physics that produces a zero or an 
additional pole in the scattering amplitude for the charm mesons near 
threshold.

We have pointed out that a triangle singularity produces a narrow peak in the cross section
 for $e^+ e^-$ annihilation into $X\gamma$ at an energy about 2.2~MeV above the $D^{*0} \bar D^{*0}$ threshold.
We gave a normalized prediction for the cross section in that region.
The peak from the triangle singularity is large enough that it could be observable by the BESIII detector.
The observation of this peak would provide strong support for the identification of the $X(3872)$ 
as a weakly bound charm-meson molecule.

\begin{acknowledgments}
This work was supported in part by the Department of Energy under Grant No. DE-SC0011726, and by the National Science Foundation under Grant No. PHY-1607190.
\end{acknowledgments}


 \end{document}